\documentclass{emulateapj}
\usepackage{natbib}
\usepackage{graphicx}
\usepackage{amssymb}

\newcommand{\be}{\begin{equation}}
\newcommand{\ba}{\begin{eqnarray}}
\newcommand{\ee}{\end{equation}}
\newcommand{\ea}{\end{eqnarray}}

\def\lesssim{\mathrel{\hbox{\rlap{\hbox{\lower4pt\hbox{$\sim$}}}\hbox{$<$}}}}
\def\gtrsim{\mathrel{\hbox{\rlap{\hbox{\lower4pt\hbox{$\sim$}}}\hbox{$>$}}}}

\def\simless{\mathbin{\lower 3pt\hbox
   {$\rlap{\raise 5pt\hbox{$\char'074$}}\mathchar''7218$}}}   
\def\simgreat{\mathbin{\lower 3pt\hbox
   {$\rlap{\raise 5pt\hbox{$\char'076$}}\mathchar''7218$}}}   
\def\apj{ApJ}
\def\apjs{ApJS}
\def\apjl{ApJL}

\def\mnras{MNRAS}

\begin{document}

\submitted{submitted to ApJL}

\title{Reconstructing the Thomson Optical Depth due to Patchy Reionization  
with 21-cm Fluctuation Maps}

\author{Gilbert P. Holder\altaffilmark{1},Ilian T. Iliev\altaffilmark{2},
Garrelt Mellema\altaffilmark{3}}
\altaffiltext{1}{Department of Physics, McGill University, Montreal, QC H3A
  2T8; Canada Research Chair; CIAR Scholar} 
\altaffiltext{2}{Canadian Institute for Theoretical Astrophysics, University
  of Toronto, 60 St. George Street, Toronto, ON M5S 3H8, Canada}
\altaffiltext{3}{Stockholm Observatory, AlbaNova
  University Center, Stockholm University, SE-106 91 Stockholm, Sweden}
\label{firstpage}

\begin{abstract}
Large fluctuations in the electron column density can occur during the
reionization process.  We investigate the possibility of deriving the
electron density fluctuations through detailed mapping of the
redshifted 21-cm emission from the neutral medium during
reionization. We find that the electron-scattering optical depth and
21-cm differential brightness temperature are strongly
anti-correlated, allowing optical depth estimates based entirely on
redshifted 21-cm measurements. This should help to isolate the 
cosmic microwave background (CMB)
polarization fluctuations due to reionization, allowing the removal of
the patchy reionization polarization signal from the other
polarization signals and a measurement of the primordial 
CMB quadrupole at
various locations in the universe at the epoch of reionization. This
latter application in principle allows three-dimensional
mapping of the primordial
density field at $z\sim 1100$ over a large fraction of the Hubble
volume.
\end{abstract}
\section{Introduction}

Two new frontiers in astrophysics and cosmology are the search for small angle
polarization anisotropies in the cosmic microwave background and the efforts
to map the redshifted 21-cm emission from neutral hydrogen at the epoch of
reionization.
On small angular scales, a significant source of CMB polarization comes from
Thomson scattering of the primordial quadrupole anisotropy by compact regions of
ionized gas. These ionized regions would appear in maps of the neutral gas as
holes in a relatively uniform map. Thus we expect that where there are strong 
sources of Thomson scattering there should be missing 21-cm emission and where
there is strong 21-cm emission there should be low Thomson optical depth, 
$\tau_{\rm es}$, i.e. the two should be anti-correlated.
In this paper we investigate the relation between $\tau_{\rm es}$ and
neutral hydrogen optical depth 
in a quantitative way that properly captures the relevant physics 
using numerical simulations of structure formation with radiative transfer.

This topic is also related to the idea of Kamionkowski \& Loeb\ (1997)
\nocite{kamionkowski97a} for using polarized CMB anisotropy due to
Thomson scattering by galaxy clusters of the CMB temperature
quadrupole as a means to extract more information about the density
field at $z\sim 1100$ than we can measure in the CMB
\citep{sazonov99,seto00,baumann03,portsmouth04,seto05,bunn06,shimon06}.
The information content of the quadrupole measurements has been
carefully investigated by Bunn (2006) \nocite{bunn06}, where it was
found that higher redshift clusters are preferred for this, since
there is less overlap with our measured CMB sky. Cluster $\tau_{\rm
es}$ are typically around 0.005 \citep{mason00}, and we will show that
typical fluctuations of $\tau_{\rm es}$ during reionization can be of
comparable size.  The quadrupole at the time of reionization should
not be affected by the integrated Sachs-Wolfe effect and thus provides
a remarkably clean measure of very large scale structure at $z\sim 1100$.

Reionization occurs in a patchy way: strongly-clustered 
sources create large ionized regions embedded in a mainly neutral medium, 
which generates CMB polarization anisotropies on small scales. Scattering of 
the primordial CMB temperature quadrupole leads to linearly polarized 
signals from ionized regions, where the size of the ionized bubble sets the 
scale of the anisotropy. On large scales, the relevant quantity is 
just the mean ionization as a function of redshift, since the small scale 
structure is averaged out. This is the signal that WMAP 
detected as the hallmark of reionization. 

The amplitude of the expected signal is expected to be on the order of the
fluctuations in the electron scattering optical depth times the amplitude
of the relevant primordial quadrupole. The latter should be on the order 
of 15 $\mu K$  but the former is not well constrained at present. 
Calculations done assuming that the ionization fluctuations can be treated 
as a Gaussian random field find very small fluctuations, since modes along 
the line of sight (LOS) tend to average down \citep{hu00}. However, recent 
simulations \citep{2006MNRAS.369.1625I,21cmreionpaper} 
 show surprisingly large fluctuations in the optical depth, on the order of 
0.01 coming just from the epoch of reionization, as we will show below. This 
optical depth is larger than that found in even the largest galaxy
clusters in the local universe. The all-or-nothing nature of the
ionized regions effectively breaks the Limber approximation, often 
leading to very little cancellation along the LOS.
Recent work has suggested that patchy reionization scenarios can result in 
interesting levels of CMB polarization \citep{mortonson06}.

In principle, one can use a H~I map to synthesize an 
effective $\tau_{\rm es}$ map and  
compare with a CMB polarization map to infer the CMB
quadrupole at the location of the scattering. Naively, the fluctuation 
maps (after subtraction of the spatial means) should differ only by an 
overall amplitude scaling, provided the primordial quadrupole is slowly 
varying over the epoch of reionization.

We assume a flat $\Lambda$CDM cosmology with parameters 
($\Omega_m,\Omega_\Lambda,\Omega_b,h,\sigma_8,n)=(0.27,0.73,0.044,0.7,0.9,1)$
\citep{2003ApJS..148..175S}, where 
$\Omega_m$, $\Omega_\Lambda$, and $\Omega_b$ are the total matter, vacuum, 
and baryonic densities in units of the critical density, $\sigma_8$ is the 
present rms linear density fluctuation on the scale of 
$8 h^{-1}{\rm Mpc}$, and $n$ is 
the index of the primordial power spectrum of density fluctuations.

\section{Simulations}

Our basic methodology was described in detail in \citet{2006MNRAS.369.1625I}.
We start by performing a high resolution $100\,h^{-1}$Mpc 
N-body simulation with a spatial 
grid of $3248^3$ cells, $1624^3=4.3$ billion particles, and
using the
particle-mesh code PMFAST \citep{2005NewA...10..393M}. This yields detailed
halo catalogues, and density and velocity fields at up to 100 roughly
equally-spaced times. All halos identified in our simulation volume are
assumed to be sources of ionizing radiation with ionizing photon emissivity
given by a constant mass-to-light ratio. We follow the time-dependent
propagation of the ionization fronts produced by sources in the simulation
volume using a detailed radiative transfer and non-equilibrium chemistry code
called $C^2$-Ray~\citep{methodpaper}, which has been extensively tested against 
available analytical solutions~\citep{methodpaper} and a number of other 
cosmological radiative transfer codes \citep{2006MNRAS.tmp..873I}. The transport
of ionizing radiation is done on a coarsened grid of $203^3$ or $406^3$ cells,
in order to make the problem tractable. These simulations allow for a number
of detailed predictions of 21-cm \citep{21cmreionpaper} and kSZ signals
\citep[][; Iliev et al. in prep.]{2006astro.ph..7209I}. In this work we use
data from simulation f2000 \citep[see][]{21cmreionpaper}, 
though our conclusions are generic.

\section{Fluctuations from Patchy Reionization} 

\subsection{21-cm Emission Fluctuations}

For our assumed cosmology (where dark energy and curvature are dynamically 
unimportant at the epoch of reionization) and ignoring the peculiar motions 
(i.e. assuming that the line profile is determined solely by the Hubble 
expansion) the 21-cm optical depth of a hydrogen cloud can be written as
\begin{eqnarray}
\tau (\nu) 
&=&  \frac{3h c^3 A_{10}}{32\pi k_B T_S \nu_0^2} \frac{n_{\rm HI}(z)}{H(z)} 
\nonumber\\
 &=& {8.5 mK \over T_S} x_{\rm HI}(1+\delta_\rho)
(1+z)^{3/2},
\end{eqnarray}
where $A_{10}=2.85\times10^{-15}$ is the Einstein A-coefficient for spin-flip 
transition, $T_S$ is the 21-cm transition spin temperature, $\nu_0$ is the 
rest-frame line frequency, $H(z)= H_0{\Omega_m}^{1/2} (1+z)^{3/2}$ is the 
Hubble constant at redshift $z$, and 
$n_{\rm HI}(z)$ is the hydrogen number density, $x_{\rm HI}$ is the mass-weighted
neutral fraction of hydrogen and $\delta_\rho$ is the local overdensity 
\citep[e.g.][]{2002ApJ...572L.123I}. 
For simplicity, we assume that the spin 
temperature is much greater than the CMB temperature, which should be a good 
approximation after the earliest phases of reionization, and that $\tau\ll1$,
in which case the 21-cm differential brightness temperature is given by:
\begin{eqnarray}
\Delta T_b(\nu) &=& {T_S -T_{\rm cmb} \over 1+z }[1-e^{-\tau(\nu)}]\nonumber\\
&\approx&  8.5\, mK \,x_{HI} \left(1+ {\delta_{\rho}}\right) (1+z)^{1/2}.
\label{dTb} 
\end{eqnarray}

We can express the neutral fraction in terms of the mean ionized fraction and
fluctuations around the mean 
\begin{equation}
x_{\rm HI} \equiv 1-\bar{x}_m-\delta x,
\label{x_HI}
\end{equation}
where $\bar{x}_m$ is the mean mass-weighted ionized fraction and
$\delta x$ is the fluctuation in the ionized hydrogen fraction. 
Combining equations~(\ref{dTb}) and (\ref{x_HI}) yields
\ba
\Delta T_b(\nu) &\approx&   8.5\, mK 
(1+z)^{1/2}  \nonumber\\
&\times&[1 - \bar{x}_{m} + (1 - \bar{x}_m) \delta_{\rho}
    - \delta x (1+\delta_{\rho})
].
\ea

If we subtract off the {\em spatial} mean, as most observing schemes
necessarily would, then the fluctuations in the 21-cm emission are given by  
\ba
\delta[\Delta T_b(\nu)] &\approx& 8.5 \, mK  
(1+z)^{1/2}  
\nonumber\\&\times&
  [(1 - \bar{x}_{m}) \delta_{\rho}
    - \delta x(1+\delta_{\rho})
].
\label{dTb_equ}
\ea

\subsection{Thomson optical depth fluctuations}

The optical depth to Thomson scattering is 
\begin{equation}
\tau_{\rm es} = \int n_e(z) \, \sigma_T \, c \, dt,
\end{equation}
where $\sigma_T$ is the Thomson cross-section.
There is some subtlety here in relating this to the hydrogen distribution,
since some of the free electrons come from ionized helium. For simplicity
 we assume that fluctuations in the helium ionization trace the hydrogen 
ionization. Extreme variations in the ionization state of helium could change
the optical depth fractionally at the 10\% level. Using $N_{\rm He}$ to denote
the number of electrons from helium ionizations per hydrogen ionization
\ba
\tau_{\rm es} &=& \int (1+N_{\rm He})  n_H(0) (1+z)^2 \sigma_T \, {c \over H(z)} 
\nonumber  \\
&\times&[\bar{x}_{m} (1+\delta_\rho) 
 + \delta x (1+ {\delta_\rho})]    dz.
\ea
Subtracting the spatially uniform part we find
\begin{eqnarray}
\delta \tau_{\rm es} 
 & = & 0.0031  (1+N_{\rm He}) 
 \nonumber  \\ &\times&
\int dz [\bar{x}_{m} {\delta_\rho}
 + \delta x (1+ {\delta_\rho})] (1+z)^{1/2}.
\label{dtau_equ}
\end{eqnarray}

\begin{figure*}
\centerline{
\includegraphics[width=2.3in]{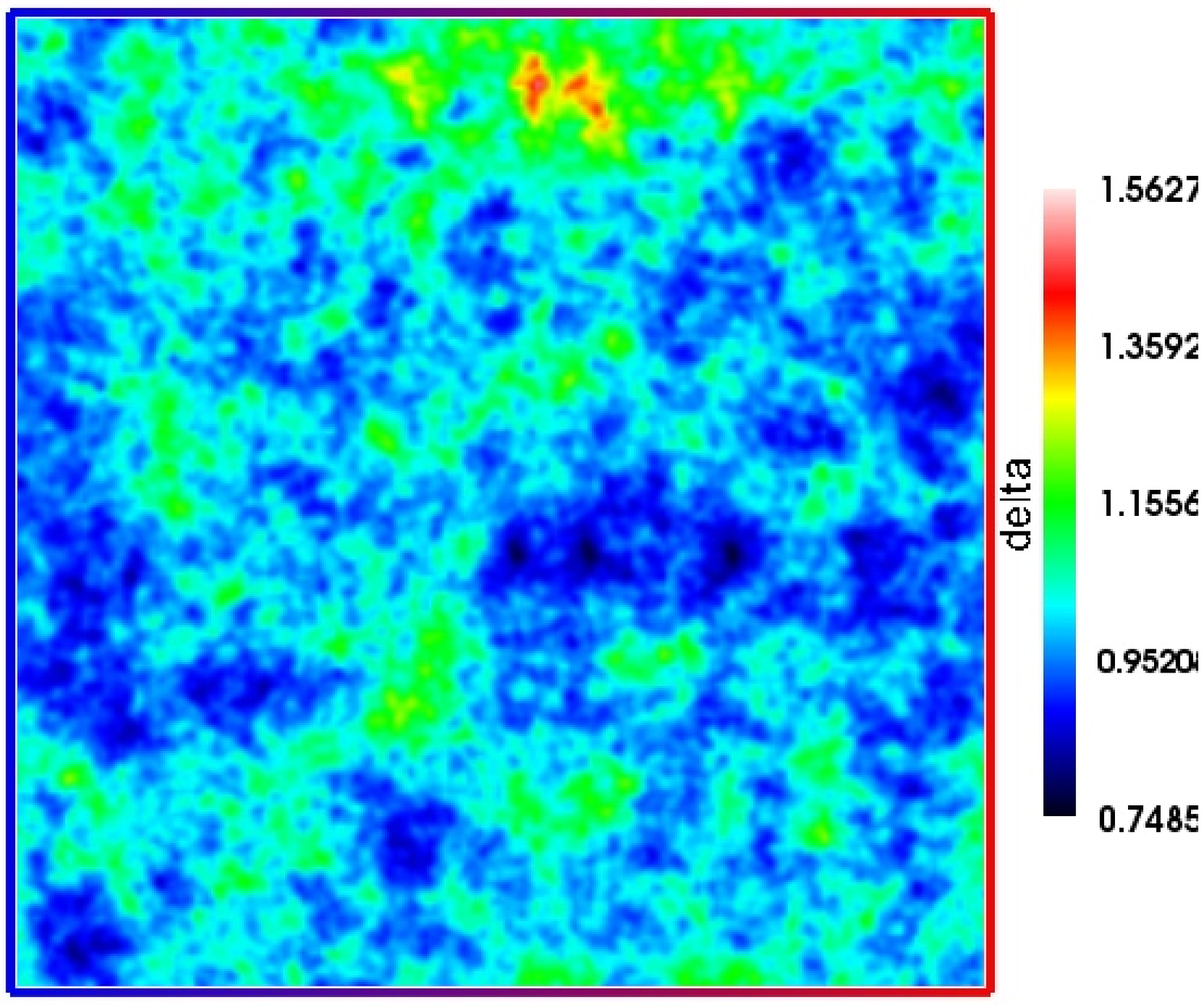}
\includegraphics[width=2.3in]{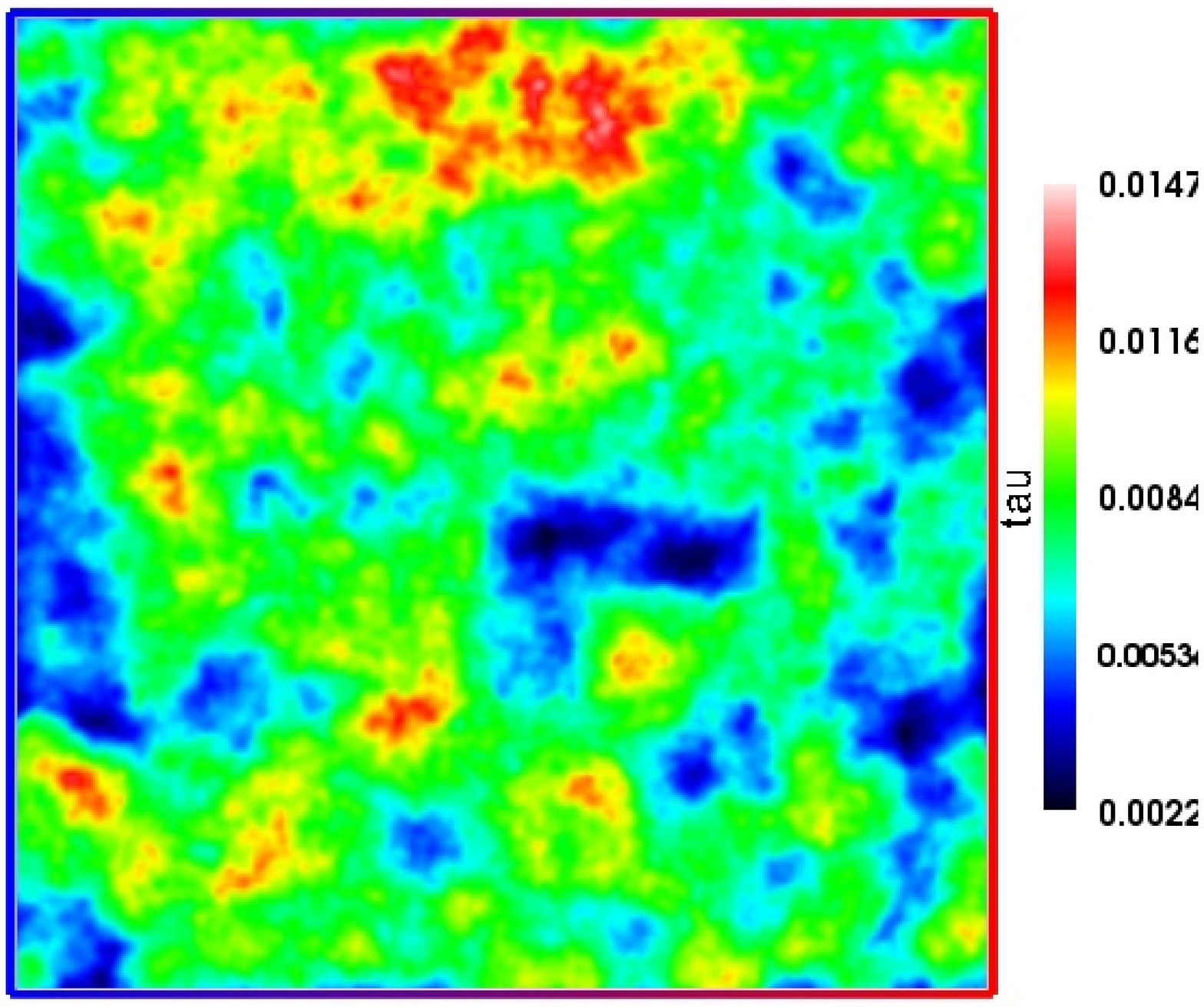}
\includegraphics[width=2.3in]{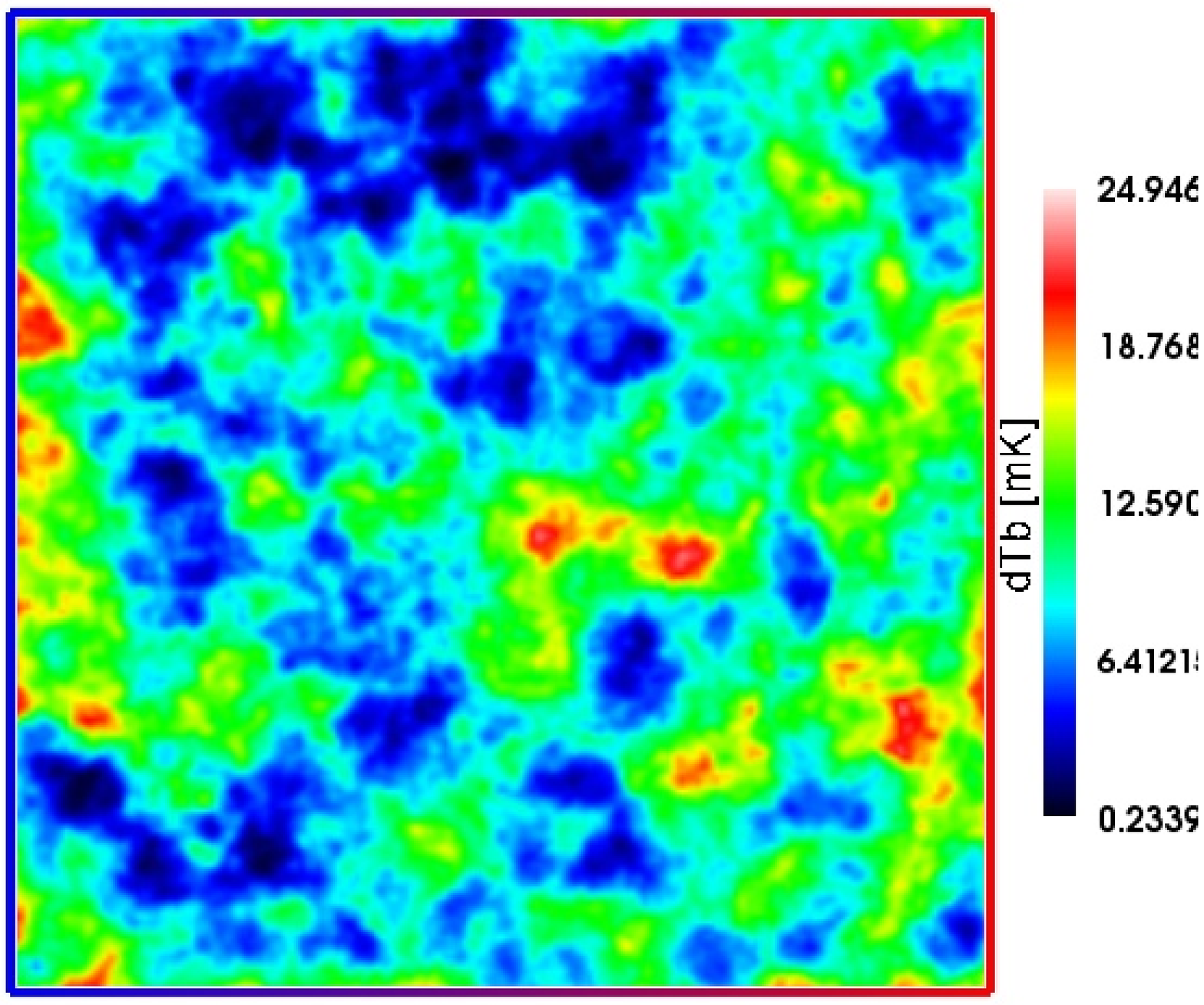}
}
\caption{Maps of a single redshift output (z=12.9) showing integrated
  overdensity (left) integrated Thomson optical depth (middle) and integrated
  21-cm emission (right) through the computational volume ($6.6$~MHz
  bandwidth). Each map is 47.5' on a side, with each pixel
  about 14''.  
  (Maps were produced using the package IFRIT of N. Gnedin). 
\label{fig:maps}}
\centerline{
\includegraphics[width=6.in]{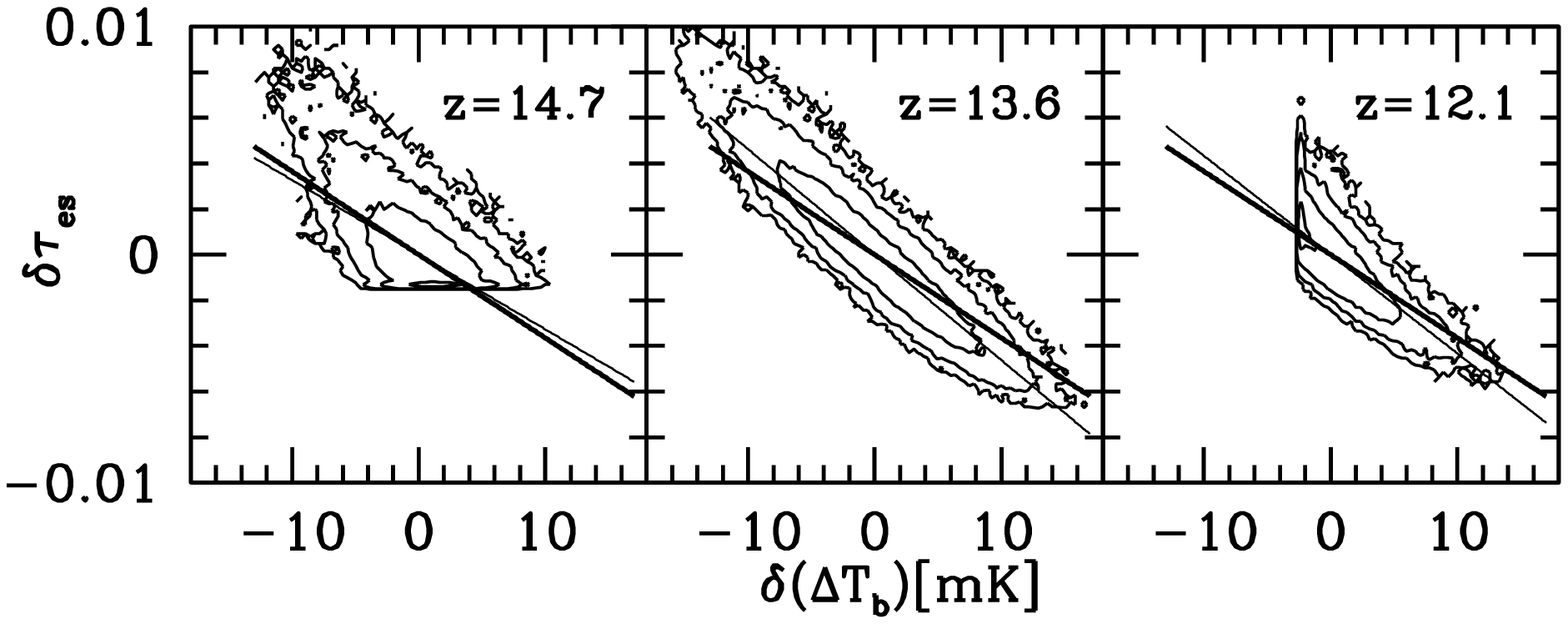}
}
\caption{Contour plots from maps constructed from single redshift output 
showing integrated optical depth vs 21-cm integrated intensity. 
Shown is the distribution of map pixels, with contours showing a density
of points that is 1, 10, and 100 points per cell.
From left to right shows redshift 11.9, 12.6 and 13.6. The heavy line
shows the expectation in the absence of fluctuations in the projected
density map. Scatter comes from density fluctuations,
while the change in slope (shown by the thin line) arises due to
density-ionization correlations.
\label{fig:tau_v_dt} }
\end{figure*}

\section{Combining data sets}

Sample maps with bandwidth of $\sim6.6$~MHz (corresponding to the whole
computational volume) of the integrated overdensity, Thomson optical depth,
and 21-cm differential brightness temperature fluctuations are shown in 
Figure~\ref{fig:maps}. If the perturbations were small, linear and Gaussian, 
$\delta \tau_{\rm es}$ would be small. 
From the maps it is apparent that this is not the case. 
The large-bandwidth 21-cm map is an excellent negative image of the
optical depth map, indicating that 21-cm imaging can in fact be used for
inferring the optical depth fluctuations. Comparing equations~(\ref{dTb_equ})
and (\ref{dtau_equ}) we find
\begin{equation}
\delta\tau_{\rm es} = 0.0031(1+ N_{\rm He})
\int  \Bigl ( 
 {\delta_\rho}(1+z)^{1/2} 
-\frac{\delta\left\{\Delta T[\nu(z)] \right\}}{8.5\,\rm mK} 
      \Bigr ) dz.
\label{theor_slope}
\end{equation}
The only differences between an actual $\tau_{\rm es}$ fluctuation map 
and a $\tau_{\rm es}$ fluctuation map derived from a 21-cm map will 
come from the first term above, i.e. from the density fluctuations along the 
the LOS, plus any nonlinear effects. This gives 
a small signal, as Figure~\ref{fig:maps} shows, since there are many effective 
volumes along the LOS giving positive and negative contributions,
 which largely cancel each other. Otherwise, the fluctuation maps will be 
identical on a pixel-by-pixel basis.

The ionization-density correlations are non-trivial. In
Figure~\ref{fig:tau_v_dt} we show a the distribution of 
$\delta\tau_{\rm es}$ vs. $\delta(\Delta T_b)$ for a 
few different redshifts. The fluctuations are strongly anti-correlated. 
Taking $N_{\rm He}=1.08$ (i.e. assuming singly ionized helium everywhere 
hydrogen is ionized as done in the simulation), and assuming that the density 
fluctuations 
average to zero, equation~(\ref{theor_slope}) gives the 
expected relation: $\delta(\Delta T_b)[K] \sim 2.54\, \delta\tau_{\rm es}$. 
The best-fit slope found in the simulations (shown with thin line in 
Figure~\ref{fig:tau_v_dt}) differs from 
the simplest expectation (thick line) at a noticeable level. The scatter arises
because the mean overdensity in the maps is not zero, as shown in
Figure~\ref{fig:maps}. Furthermore, these density fluctuations are correlated 
with optical depth and 21-cm emission. Physically this results from the inside-out,
rather than random nature of the ionization process. The denser regions are on 
average ionized earlier, and are also more non-linear (as evidenced by 
the larger deviations in the upper left of each curve). 
The best fit slope varies with redshift, but within this 
ionization fraction range is never different from the theoretical value by more 
than $25\%$. This suggests that even with no understanding of the
reionization process it will be possible to use 21-cm maps to reconstruct
$\tau_{\rm es}$ to better than 25\% accuracy. From the morphology of the 21-cm 
fluctuations, it should be possible to understand the reionization process at 
a level that allows a much better understanding of this slope.

The 21-cm fluctuations can thus be used to 
reconstruct the optical depth fluctuation map that leads to CMB polarization.
With a CMB polarization map in hand of sufficient quality one could 
simply do a direct template search to determine the CMB quadrupole at
a given location at the epoch of reionization. 
Residual contamination will come from velocity-induced quadrupoles during 
reionization, which are expected to be roughly  10\% of the primordial
quadrupole.

\section{Observational Prospects}

The signals discussed here are small. The 21-cm emission fluctuations on 
scales of order tens of arcseconds are of order a few mK. For wavelengths 
of several meters the diffraction limit corresponding to resolution of tens 
of arcseconds requires effective aperture sizes of tens of km. The LOS extent in
frequency space is of order 1 MHz, and the system temperature is $\sim$500K, 
set by the Galaxy \citep[e.g.][]{furlanetto06}. For a single dish big
enough to just resolve the sources (tens of km across), the
noise scales as $T_{sys}/\sqrt{\Delta \nu \Delta t}$ for bandwidth $\Delta
\nu$ and observing time $\Delta t$. It would take roughly an hour to integrate
down to 5 mK with this instrument. Using a smaller aperture dilutes the signal
by the square of the telescope diameter, $d$, and thus increases the requisite
integration time by $d^4$. This will be a challenging signal that
will require square kilometers of collecting area. 

Imaging the CMB polarization on scales of tens of arcseconds to the requisite
sensitivity will not be easy. The signal will be on the order of the
primordial quadrupole times the Thomson optical depth, or roughly 0.1 $\mu
K$. The system temperatures at mm wavelengths are tens of
Kelvin at best. A single telescope 
operating at mm wavelengths with a 20 GHz bandwidth and a system
temperature of 50K would require roughly six months of integration time to 
reach 0.1 $\mu K$ sensitivity per diffraction-limited beam element. At
2mm, a 10'' beam requires a 40m aperture. The Atacama Large Millimeter Array
(ALMA) in a compact configuration roughly has this angular
resolution and collecting area, with slightly more collecting area
than is required, but an array filling factor that will resolve out some of
the flux.  A rough estimate of ALMA observing times yields time scales of
months of continuous integration. If giant ionized regions exist around large 
clusters of sources, then this could lead to
signals that are an order of magnitude larger and all integration times thus
drop by two orders of magnitude. If such larger ionized regions are visible in
21-cm brightness fluctuations then ALMA will be able to image the CMB
polarization directly. However, imaging the general morphology of reionization
in 21-cm emission and CMB polarization will both be challenging undertakings.

\section{Summary and Discussion}

We have shown that fluctuations in 21-cm emission from the epoch of
reionization and CMB polarization fluctuations should trace
each other extremely well. High quality redshifted 21-cm
observations will be able to reconstruct the optical depth to Thomson
scattering to better than $20\%$ accuracy; better accuracy can
be achieved with a better understanding of reionization.

A map of the Thomson optical depth obtained from 21-cm emission will allow at
least two important improvements in our knowledge. If
reionization is a contributor to foreground ``B mode'' polarization
anisotropy, this will allow cleaning of this foreground. Also, 
the polarization signal can be used to measure the primordial
quadrupole at the time of reionization. A
coarse grid of all-sky coverage will allow reconstruction of
the large scale structure in the universe; 
this includes the volume of the Dark Ages, 
which is currently unobservable due to the lack of sources. With 
dense coverage a reconstruction of nearly the entire primordial 
density field at a snapshot of $z\sim 1100$ within
our Hubble volume would be possible; since each point only provides the
quadrupole (a very large scale convolution) it is not clear that the
reconstruction would be accurate in the presence of astrophysical
contaminants, noise, and bulk flows that contribute velocity
quadrupoles. However, this signal contains {\em unique} information
about the density fluctuations at $z \sim 1100$ over a large
fraction of our Hubble volume.

Thomson scattering of the primordial quadrupole at lower redshifts has been
discussed in the context of using galaxy clusters as Thomson scatterers, but
the 21-cm emission has several important advantages: 1) the
Thomson optical depth can be accurately reconstructed from the 21-cm map, 
while galaxy cluster optical depths may not be measurable at the
requisite sensitivity; the optical depth derived from Sunyaev-Zel'dovich
measurements is weighted by the Compton $y$-parameter and is not
appropriate for this purpose \citep{knox04}; 
2) the overlap in information of the CMB
quadrupole at $z\sim 10$ and  our measured CMB sky is much reduced when
compared with clusters at $z \sim 0.2-0.5$; local clusters are better
as probes of the ISW effect; 
3) the Sunyaev-Zel'dovich effect and gravitational lensing
of the CMB will not be correlated with the CMB polarization signal
and one does not expect radio halos, quasar and galaxy overdensities, or
strongly lensed background objects.

With only a polarization map one could still measure the direction of the CMB
quadrupole, which would allow tests of isotropy and homogeneity of the CMB at
$z\sim 10$, provided the reionization process is relatively sharp in
redshift. Large ionized regions around quasars would provide interesting
targets for ALMA to try to measure this signal, although imaging a field to
$\mu K$ sensitivity with a quasar at the center will require some care.
While the signals will be difficult to image with high signal to noise, these 
observations offer tremendous opportunities for gaining information about our 
universe that is not available by any other means. The hurdles are not
insurmountable and the observations required (high signal to noise neutral 
hydrogen maps and high resolution imaging of the CMB polarization fluctuations)
are already on the ultimate wish list for physical cosmology.

\acknowledgements{\vskip -0.30in GPH is supported by an NSERC Discovery grant. We
thank Olivier Dore and Ue-Li Pen for useful discussions.}




\end{document}